\documentclass[11pt,letterpaper]{article}

\usepackage{paper}
\usepackage{natbib}
\usepackage[colorlinks=true,linkcolor=blue!60!black,citecolor=blue!60!black,urlcolor=blue!60!black]{hyperref}

\usepackage{algorithm}
\usepackage{algorithmic}
\usepackage{newfloat}
\usepackage{listings}
\DeclareCaptionStyle{ruled}{labelfont=normalfont,labelsep=colon,strut=off}
\floatstyle{ruled}
\newfloat{listing}{tb}{lst}{}
\floatname{listing}{Listing}

\usepackage{tikz}
\usepackage{pgfplots}
\pgfplotsset{compat=1.18}
\newcommand{\inlineheading}[1]{%
  \par\vspace{0.5em}%
  \noindent\textbf{#1.}\ %
}

\title{Qualifying and Quantifying Risk Under the EU AI Act}

\author[1]{Gustavo Gil Gasiola}
\author[2]{Sarah H. Cen}
\author[1,3]{Frederike Zufall}
\affil[1]{Department of Informatics, Karlsruhe Institute of Technology (KIT)}
\affil[2]{Departments of Electrical \& Computer Engineering and Engineering \& Public Policy,\newline
Carnegie Mellon University}
\affil[3]{Waseda University}

\date{}

\begin{document}

\maketitle
\begingroup
\renewcommand\thefootnote{}
\footnotetext{Correspondence to: \texttt{gustavo.gasiola@kit.edu}, \texttt{sarahcen@andrew.cmu.edu}, \texttt{zufall@kit.edu}.}
\endgroup
\setcounter{footnote}{0}

\begin{abstract}
The EU AI Act uses a risk-based approach to regulate AI systems, calibrating the intensity of regulation according to the risks they pose. While the term `risk' implies quantification, resulting from the combination of the probability and severity of harm, the AI Act refers to risks to fundamental rights, thereby engaging a qualitative perspective. In this piece, we address this puzzle using a two-step framework under which the EU AI Act balances risks with the protection of fundamental rights, the legitimate purposes of providers and deployers, and the impacts of regulatory measures on providers, deployers, and regulators. We discuss this framework against the backdrop of potential approaches to quantifying risks, with a specific focus on defining and measuring the main components of the concept of risk: probability, severity, and their combination. We suggest that the protection of fundamental rights and risk quantification can be aligned by incorporating quantification methodologies into the proposed framework. In particular, the AI Act implies a balancing analysis that uses a severity-first approach to classify and quantify the risks posed by AI systems. 
The integrated framework not only helps to clarify the AI Act's risk-based approach, but can also inform technical and implementation choices. Finally, we conclude that if the risk quantification methodology or its application to the protection of fundamental rights is left to providers and deployers, there is potential for `risk hacking', which could lead to the underclassification of AI systems and subsequent regulatory shortcuts.

\end{abstract}

\section{Introduction and Motivation}

The EU AI Act\footnote{Regulation (EU) 2024/1689 of the European Parliament and of the Council of 13 June 2024  laying down harmonised rules on artificial intelligence and amending Regulations (EC) No 300/2008, (EU) No 167/2013, (EU) No 168/2013, (EU) 2018/858, (EU) 2018/1139 and (EU) 2019/2144 and Directives 2014/90/EU, (EU) 2016/797 and (EU) 2020/1828 (Artificial Intelligence Act).} (AIA) categorises and regulates AI systems using a `risk-based approach' (Recital 26 AIA). The concept of risk is used to both define the applicable rules \citep{gellert2021role, koulu2023artificial, gasiola2025rebuilding} and to calibrate compliance with the mandatory requirements \citep{degregorio2022european}. 
Broadly, the AI Act divides AI systems into three risk levels: unacceptable risk, high risk, and minimal risk.\footnote{We do not include the limited risk classification \citep{comm2024aiactquestions}, as legal scholars have argued that it is not a distinct risk level \citep{gasiola2025rebuilding}.}
Systems with unacceptable risk are prohibited, those with high or limited risk are subject to mandatory requirements, and the remainder are left unregulated.
Thus, risk is central to the AI Act: obligations scale with the assessed risk of an AI system, 
and, consequently, providers and deployers are incentivised to seek classification of their systems at lower risk levels.
For instance, Art. 6(3) AIA allows providers and deployers of AI systems used in the critical areas listed in Annex III to avoid classification as high risk if they deem their systems to pose no significant risk of harm. Even after classification, the quantification of risks remains relevant. For high-risk AI systems, Art. 9 AIA requires the implementation of risk management systems. This encompasses the implementation of risk mitigation measures defined by providers and deployers \citep{degregorio2022european}. 
Under Art. 7 AIA, the Commission also needs to quantify risks when adding or modifying high-risk use cases listed in Annex III. 
This evaluation of risk is a continuous process involving identifying and estimating risks as well as adopting appropriate measures to minimise them below an acceptable threshold. 

Yet the meaning of risk is not immediately clear.
Art.~3(2)~AIA defines risk as `the combination of the probability of an occurrence of harm and the severity of that harm'. 
At the same time, the Act establishes that `risk' should be understood as harms to `health, safety or fundamental rights'.\footnote{See in particular Art. 6(3) and 9(2)(a) AIA.}
The first definition already leaves room for interpretation of the terms `probability', `severity', and even `combination' \citep{feiler2025article3}.
In addition, it is unclear whether the definition is compatible with the concept of `risk' as harms to `health, safety or fundamental rights'.
The definition implies that risk can be quantified, but applying a quantitative analysis to qualitative concepts may not be possible.
In particular, one could argue that the exercise of measuring the value of a fundamental right is inherently contradictory \citep{gellert2021role, yeung2022demystifying, malgieri2025assessing}. 

Operationalising the notion of `risk' is necessary for the effective implementation and enforcement of the AI Act.
Thus, the Commission, when implementing guidelines, would have to seriously consider how to interpret `probability', `severity', and their `combination' and reconcile what it means to evaluate risk in the context of fundamental rights.
Aligning the presumably quantitative definition of risk with qualitative legal concepts is also relevant in light of the competing incentives: regulators may wish to classify a system as high-risk  to subject it to greater scrutiny, but providers and deployers would likely resist this and could potentially even `manipulate' their measurable features to avoid such classification. 
Consensus and clarity on the meaning of risk can mitigate the chances of this `risk hacking' outcome.

In this piece, we first interpret the AI Act's risk-based approach through a two-step framework that balances risk with the protection of fundamental rights, the legitimate purposes of providers and deployers, and the impact of regulatory measures. We then examine potential approaches to quantifying risks, with a specific focus on defining and measuring probability, severity, and their combination. Finally, we argue that, while the chosen methodology is political in nature, not all risk quantification approaches are compatible with safeguarding fundamental rights in a risk-based regulatory context.

\section{Dimensions of Risk in the EU AI Act}\label{sec:dimensions}

In this section, we review the role of `risk' in EU law and, in particular, in the AI Act.
While it may seem that the protection of fundamental rights is incompatible with a quantitative operationalisation of the definition of `risk' (Art.~3(2)~AIA), we show that the AI Act's risk classification is a balancing decision that weighs an AI system's potential harm to fundamental rights, health, and safety alongside the system's intended purpose.
In order to keep risks under an acceptable threshold, its risk classification also considers the impact of regulatory measures --– regarded as risk management strategies --– on providers, deployers, and regulators.

\subsection{A Reciprocal Relationship Between Risk and Regulation}

In EU law, the concept of risk first appeared in regulations aimed at controlling environmental, health, and safety hazards, followed by new technologies and industries \citep{macenaite2017riskification}. In these areas, the so-called precautionary principle
played a key role in justifying preventive measures in cases of scientific uncertainty based on risk assessment\footnote{Cf. CJEU, \textit{Alpharma Inc. v. Council of the EU}, Case T-70/99, par. 135.} \citep{commission2000communication, stirling2017precaution}. Over time, the use of risk as a regulatory technique has grown considerably, particularly for AI systems \citep{kaminski2023regulating} and within the Digital Market Strategy \citep{degregorio2022european, macenaite2017riskification} to address novel risks posed by technology. 

The concept of risk has two main functions in regulatory frameworks: while `risk regulation' uses risk as the rationale behind or the object of regulatory intervention \citep{black2010therole, degregorio2022european}, a risk-based approach relies on risk scores to calibrate regulation \citep{baldwin2011understanding,quelle2018enhancing} and guarantee proportionality \citep{gasiola2025rebuilding}. Despite these counter-referential relationships between `risk' and `regulation', both functions can be incorporated into the same legal instrument, to regulate risks using a risk-based approach \citep{degregorio2022european}. And indeed, this combination mirrors the structure of the AI Act.

\subsection{The System of Risk in the AI Act} \label{sec:risk-based}

Art. 1(1) AIA states its objective is to ensure a high level of protection of health, safety, fundamental rights, including democracy, the rule of law, and environmental protection, against the harmful effects of AI systems. 
Within this, the so-called risk-based approach (Recital 26 AIA) classifies AI systems according to the risk they pose to the regulatory goals into three top-down defined risk levels: unacceptable, high, and minimal risk.

\textbf{Unacceptable-risk AI systems} pose excessive risks \citep{mahler2021between, burgstaller2023use, malgieri2024licensing} and are prohibited. We can find the classification criteria in Art. 5(1) AIA, which describes aspects, usages, and purposes \citep{neuwirth2023prohibited} of such AI systems. However, the AI Act partially places the responsibility for assessing the concrete risk on the provider or deployer \citep{commission2025guideline}, as the criteria are often abstract and presuppose additional evaluation.\footnote{For example, Art. 5(1)(b) AIA states that an AI system that exploits vulnerabilities of a natural person or a specific group is prohibited if it `causes or is reasonably likely to cause that person or another person significant harm'. In order to classify an AI system under this criterion, it is necessary to assess the concrete probability of significant harm caused to individuals.} Even when the system poses excessive risks, the AI Act recognizes a few concrete exceptions to classification in Art. 5(1), (2), (3), and (5), lifting the prohibition of such systems under certain conditions or for specific purposes.\footnote{For example, AI systems that infer emotion in the workplace are generally prohibited under Art. 5(1)(f) AIA, but are permitted for medical or security reasons.} These exceptions are justified by balancing the potential harmful impact on fundamental rights and protected values against the realisation of important public interest or other fundamental rights at stake \citep{gasiola2025rebuilding}.

\textbf{High-risk AI systems} pose significant risks and are subject to mandatory requirements (Recitals 7 and 46 AIA). The AI Act establishes two classification criteria for the high-risk level. The first criterion covers products that are already regulated by the EU product safety regulations (Art.~6(1)(a)~AIA). The second comprises AI systems that fall within a list of critical domains, as well as the associated use cases, which describe the purpose, the AI capability, deployer, and subject \citep{golpayegani2023high} (Art. 6(1)(b), Annex~III AIA).\footnote{For example, point 7 of Annex III lists the domain `Administration of Justice' and includes `assistance of judicial authorities in interpreting the law' as a use case.} While the Commission has the power to modify the list by adding or modifying use cases (Art. 7(1) AIA), this requires a risk assessment to determine whether a new use case poses a risk equivalent to, or greater than, that posed by the high-risk AI systems already referred to in Annex III (Art. 7(1)(b) AIA).

Classification according to Art. 6(1)(b) AIA also requires the provider or deployer to carry out a concrete risk assessment. While the AI systems that fall under Annex III are initially considered high risk, Art.~6(3)~AIA allows providers and deployers to declassify \citep{gasiola2025rebuilding} their systems when they pose no significant risks in the concrete application. Some conditions that indicate the absence of significant risks can be found in Art. 6(3)(a) to (d) AIA, although no general criteria are defined.

To be lawfully put into the European market, high-risk AI systems must comply with mandatory requirements, which aim to reduce the risks to an acceptable level \citep{fraser2024acceptable, schuett2024risk}. These requirements are further scaled according to the intended purpose and the concrete risks posed by the AI system (Art. 8(1) AIA). In particular, providers and deployers must assess the risks (Art.~9(2)(a), (b), and (c) AIA) and take appropriate measures (Art.~9(2)(d) AIA) to ensure that the residual risk remains at an acceptable level (Art. 9(5) AIA) \citep{schuett2024risk}.

At the lowest level, \textbf{minimal-risk AI systems} do not pose significant risks and are therefore not subject to mandatory requirements. This risk level is a residual category \citep{de2022humpty} that includes all AI systems not classified into the other risk levels \citep{chamberlain2023risk}. For these systems, the AI Act merely encourages the adoption of codes of conduct and voluntary compliance with the high-risk requirements (Art.~95(1) AIA).
Given the potential benefits offered by minimal-risk AI systems \citep{chamberlain2023risk}, any risks posed by them are deemed to be at an acceptable level, meaning that regulatory measures are not justified or necessary.

Overall, the risk-based approach of the AI Act is rooted in an abstract risk assessment \citep{mahler2021between, gasiola2025rebuilding} conducted by the EU legislator and the Commission to define risk levels and establish corresponding classification criteria. Once the AI Act becomes fully applicable, the risks posed by AI systems will still need to be assessed and quantified: by the Commission when updating Annex III, and by providers and deployers when classifying their systems and complying with the mandatory requirements. Yet, from the AI Act's Impact Assessment \citep{commission2021impact, mahler2021between} to the recent guidelines issued by the Commission \citep{commission2025guideline}, no specific methodology for risk calculation has been presented. 

\subsection{The Notion of `Risk': `Harm' as a Connecting Concept Between Qualitative Legal Values and Quantification}
Before investigating to what degree the legal definition of risk in Art. 3(2) AIA actually lends itself to quantification (Section \ref{sec:quant_risk}), a preliminary question of legal interpretation is: what harms does the definition of risk refer to?
The AI Act defines risk as `the combination of the probability of an occurrence of harm and the severity of that harm' (Art. 3(2)). The wording and conceptual idea behind this definition can be traced back to a technical approach to risk \citep{mahler2021between} which presupposes its calculation or quantification.\footnote{Beyond the AI Act, even though they do not define `risk' directly, both the General Data Protection Regulation (GDPR, Art.~24(1), Art. 25(1), Art. 32(1), and Recitals 75, 76, 77, 90.) and the Digital Services Act (DSA, Art. 34(1), and Recital 79) refer to the likelihood and severity of the risks to the rights and freedoms of natural persons.
A similar definition can also be found in the New Legislative Framework, e.g., Art. 3(18) Regulation (EU) 2019/1020.}    
\textbf{However, this quantitative formula is contrasted with the reference to harm as a qualitative idea:} the AI Act addresses the harmful effects of AI systems on `health, safety, and fundamental rights'.\footnote{This triad is used repeatedly in the Regulation to refer to the risks of AI systems: Recitals 1, 7, 8, 20, 46, 52, 53, 65, 66, 67, 72, 140, 157, 171, 176, Art. 6(3), and (6), Art. 7(1)(b), and (3)(a), Art. 9(2)(a), Art. 13(3)(b)(iii), Art. 14(2), Art. 36(7)(e), (8)(a), and (9)(a), Art. 57(6) and (11), Art. 79(1), Art. 82(1), and Annex IV (3)~AIA.}
While health\footnote{Health protection includes physical, psychological, and public health (Recitals 5, 29, and 110 AIA).} and safety
\footnote{Safety includes product safety (Recitals 9, 47, 50, 51, 52, 55, 87, 125, and Annex III (2) AIA), safety at work (Recital 9 AIA), public safety (Recitals 19, 33, and Art. 5(1)(h)(ii) AIA), technical robustness and safety (Recital 27 AIA), physical safety, and safety of persons or property (Recitals 33, 55, Art. 3(14), Art. 5(1)(h)(ii), and Art. 10(2)(f)~AIA).}
appear as more concrete or quantifiable targets, the reference to fundamental rights opens the definition to more abstract legal concepts. 
Fundamental rights include the rights enshrined in the Charter of Fundamental Rights of the EU (CFR), such as the right to human dignity (Art. 1), the protection of personal data (Art. 8), freedom of expression and information (Art.~11), or the right to non-discrimination (Art. 21), but also the right to an effective remedy and to a fair trial (Art.~47), or the right to good administration (Art.~41 and Recital~48 AIA). 

But even beyond that, under the triad of `health, safety, and fundamental rights', we must also consider other protected interests and values. In particular, the AI Act explicitly refers to the potential risks of AI systems to democracy,
\footnote{Including democratic process, democratic values, and democratic control: Recitals 1, 2, 8, 20, 27, 28, 56, 61, 62, 110, 120, 136, 159, 176, Art. 1(1), Art. 74(8), and Annex III (8) AIA.}
the rule of law,
\footnote{Recitals 1, 2, 8, 28, 61, 62, 176, and Art. 1(1) AIA.} 
environmental protection,
\footnote{Recitals 1, 2, 8, 27, 48, 130, 155, 176, Art. 1, Art. 3(49), Art.~46, Art. 95(2)(b), and Art. 112(7) AIA.} 
and public interest.
\footnote{Recitals 5, 7, 8, 46, and Art. 82(1) AIA. However, Recital 7 refers particularly to `public interests as regards health, safety, and fundamental rights'.} Accordingly, the notion of risk of harm in the AI Act, while considering a plethora of protected rights and values, bridges qualitative legal values to quantification approaches.

\subsection{The Risk-Based Approach Balances Harms, Purposes, and the Impacts of Regulatory Measures}\label{sec:balancing}
\begin{figure*}[t]
    \centering
    \includegraphics[width=1\textwidth]{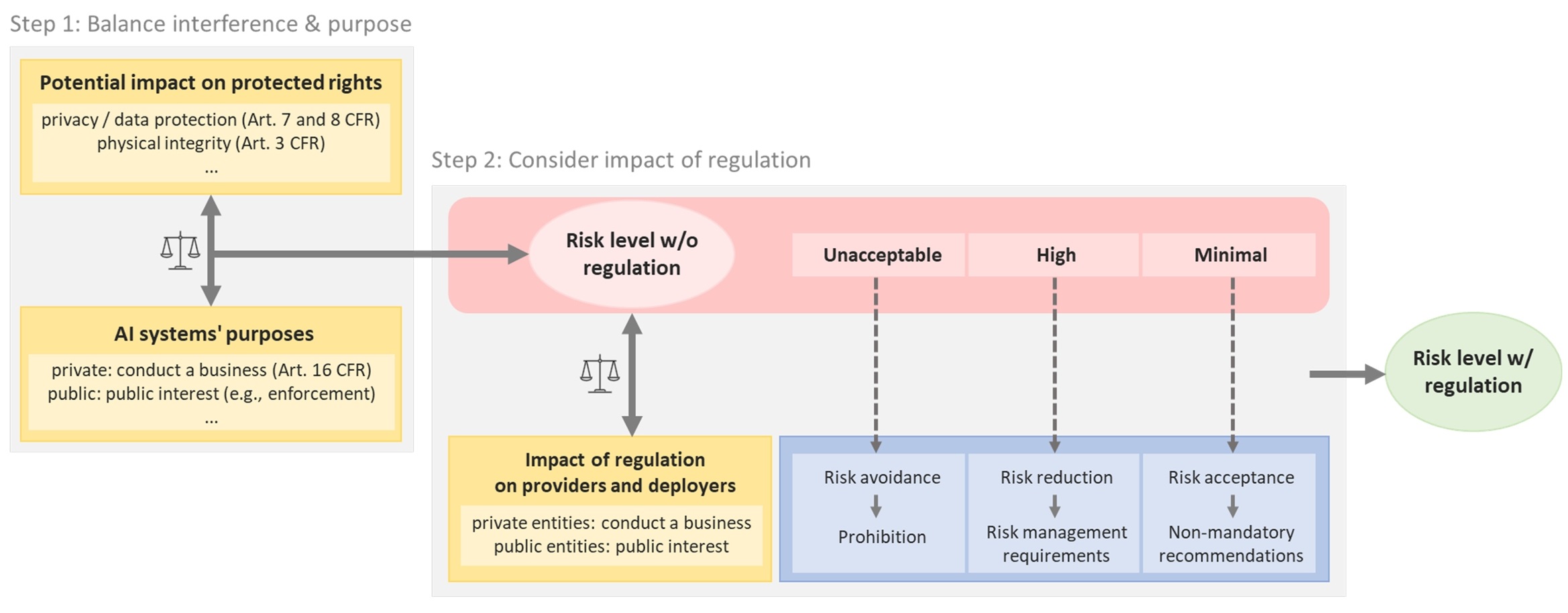}
    \caption{Two-step framework of the AI Act's risk-based approach from a fundamental rights perspective.}
    \label{fig:balancingrisks}
\end{figure*}
\textit{How can we align the doctrinal background of fundamental rights assessment with the concept of risk in the AI Act that approaches `risk' both as a quantitative formula, and in its qualitative dimension as harm to fundamental rights?} 
Before we discuss the quantification of risk in the following section, we must first reconcile quantification with the consideration of fundamental rights. 
We argue that these two concepts are not at odds, 
suggesting a two-step framework for interpreting the EU AI Act's risk-based approach (Fig.~\ref{fig:balancingrisks}).

\subsubsection{Step 1: Considering Potential Harm to Fundamental Rights in a System's Purpose} The AI Act aims to balance the protection of health, safety, and fundamental rights relative to the purposes pursued by the providers or deployers of a respective AI system, while also considering the benefits of technological innovation (Recitals 4, 5, and 6 AIA) \citep{degregorio2022european, koulu2023artificial}. 
The AI Act thus involves balancing the protection of the fundamental rights of individuals affected by AI systems with the pursuit of legitimate purposes justified by either the public interest or competing fundamental rights, including the freedom to conduct business of private entities (Art. 16 CFR). Legal theoretical and interdisciplinary research has already suggested different approaches to formalising similar balancing decisions \citep{alexy_balancing_2003,zufall2023towards,novelli2024ai,araszkiewicz_tipping_2026}. In the context of the risk-based approach, however, the outcome of the balancing act is the anticipated risk posed by the system without any regulatory intervention. 

This consideration of the intended use or purpose (rather than a presumption that the protection of fundamental rights outweighs all other considerations) is supported by the classification process set out in the AI Act, in particular the exceptions to the classification criteria. 
For instance, due to the excessive risk to fundamental rights, the deployment of emotion recognition systems in the workplace is prohibited (Art.~5(1)(f)~AIA). However, the same legal provision allows for their deployment for medical or security purposes.
Although the potential interference in this case remains excessive, the legitimate purposes of ensuring physical integrity or national security justify its classification as high risk rather than as unacceptable  \citep{gasiola2025rebuilding}. 

\subsubsection{Step 2: Managing Risks While Considering the Impact of Regulatory Interventions} 
Step 1 determines the anticipated risk based on a balancing decision between the protected fundamental rights and the interests standing behind an AI system's purpose.
Step 2 reflects the proportionate approach \citep{Coglianese2020thelaw} the AI Act takes to 
balance the risk reduction with the impact on providers, deployers, and regulators \citep{ebers2025truly} through different risk management strategies: \textbf{1) risk avoidance}: unacceptable-risk AI systems are prohibited (Art. 5 AIA) in order to avoid exposure \citep{ISO31073} to the excessive risks; \textbf{2) risk reduction}: high-risk AI systems are subject to mandatory requirements to control the risks posed (in particular, through the implementation of risk management systems, Art. 9 AIA); and \textbf{3) risk acceptance}: minimal-risk AI systems are left unregulated, as the impacts of regulation outweigh its benefits. 
In this way, the EU legislator, based on legislative discretion, prescribes regulatory interventions to lower the risk level when they outweigh the impacts on
the right to conduct business of private providers or deployers, or pose financial and organisational burdens on public entities (compare Fig.~\ref{fig:balancingrisks}). The accepted risk, in this sense, corresponds to the residual risk after the regulatory intervention.

Thus, this framework shows that the EU AI Act employs a balancing decision that considers multiple factors when classifying an AI system's risk level. 
Balancing these factors in a consistent way while adhering to the definition of risk in Art. 3(2) AIA is the subject of the next section.

\section{Quantifying Risk}\label{sec:quant_risk}

Although, based on the definition in Art. 3(2) AIA, only three components contribute to the definition of risk, there is significant complexity involved in quantifying and measuring risk. 
In this section, we discuss quantitative interpretations of each component and the corresponding implications.
Ultimately, we show that ambiguity in interpreting the three main components (`probability,' `severity,' and their `combination') can lead to different outcomes, especially `combination'. 

\subsection{Component 1: Interpreting `Probability'}\label{sec:quant_probability}

Probability is a well-accepted and well-studied term. 
Despite its popular use, probability has remained an ambiguous concept in practice. 
In this section, we discuss two considerations when interpreting probability in the wild: 
(i) determining the `\emph{events}' over which probability is defined, 
and
(ii) incorporating probability when it is difficult to \emph{measure}. 

\inlineheading{Defining the Event Over Which Probability Is Measured} When a weatherman reports that the chance of rain is 40 percent, there is ambiguity in whether this implies:
(1) under similar conditions in the past, it has rained at some point during the day approximately 40 percent of the days for which the weatherman has historical data; or 
(2) at \emph{any} point during the day, the weatherman's expectation that it will rain is 40 percent,
which implies that it will rain approximately 40 percent of the day.

These two interpretations have vastly different implications. 
For example, while the first implies that the chance that it rains by the end of the day is 40 percent, 
the second actually implies that the chance it rains by the end of the day is very high (close to 100 percent), \emph{even though both result from a forecast predicting rain with 40 percent `probability'} \citep{dawid2017individual}.
There are yet \emph{more} interpretations, 
such as that it will \emph{definitely} rain by the end of the day, but that the uncertainty is over \emph{where} it will rain, which is expected to be approximately 40 percent of the geographic area over which the weatherman is reporting.

These differences result from different notions of the `event' over which probability is measured. 
In the first setting we considered, the event is `\texttt{rains by the end of the day}' whereas the second considers the event `\texttt{rains during minute $m$ of the day}', 
and the third is `\texttt{rains at a given location}'.

In the context of the AI Act, an AI system may seemingly lead to harm with low probability under some definitions of events and high probability under others. 
For instance, suppose that the event is `\texttt{AI interaction $i$ causes harm}'.
Then, `probability' means `the chance that interaction $i$ causes harm.'
In most cases, the chance that a given interaction yields harm is low. 
However, if the event is defined as `\texttt{some AI interaction causes harm during the year}', then the probability is much higher, as it is accumulated (i) across individuals interacting with the AI system and (ii) across time.

One version of this discrepancy amounts to the \emph{difference between a low-severity event occurring frequently over the population being categorised as low-risk because the probability is taken over single interactions or high-risk due to the accumulation of harmful interactions.}
For example, online platforms can cause low-severity harms (e.g., each post on social media can affect a user's mental health in a small way), and one could argue that the probability of such events is small if they are interspersed among many other events (e.g., many other posts), but the accumulation results in potentially high risks (e.g., with respect to mental health). 

\inlineheading{Measurement Is Difficult}
Even when there is consensus on the event of interest, measuring the probability of an event can be difficult if measurement is \emph{costly} and/or the event of concern is \emph{non-repeatable}.

In particular, measuring probability may require assessing billions of pieces of data (e.g., user logs), extensively testing the system (e.g., running simulations), and tracing harms far downstream (e.g., estimating the probability that a resume-scoring tool causes broader societal harm). 
As one may expect, each of these challenges adds significant costs, which are especially difficult for small- and medium-sized companies to absorb. 
In addition, some events are so rare or forward-looking that their probability cannot be measured because the event itself has not occurred yet or occurs so infrequently. 
Unfortunately, many consequential events fall precisely into this category (e.g., the use of an AI system to create a bioweapon, or the ability of an AI system to cause self-harm), making them especially difficult to quantify. 

Measurement challenges create what is known as `epistemic uncertainty'.

While probability itself is designed to quantify uncertainty (a different type known as `aleatoric uncertainty'), there is disagreement within the community on how to incorporate \emph{epistemic} uncertainty into the formulation of probability.

For instance, a \emph{frequentist} may argue that if an event has not yet occurred in the collected data, then it has probability $0$ while a \emph{Bayesian} would use a prior over an event's probability, resulting in a non-zero probability estimate even if the event has never been observed.
Even further, there is a whole range of ways of incorporating probability into risk estimation depending on how `optimistic' or `pessimistic' one wishes to be.
For instance, suppose the goal is to estimate the risk of causing a user to self-harm;
one can err on the side of caution by using the upper confidence bound or upper quantile of a harm's probability, thereby intentionally overestimating the risk in the face of uncertainty. 
Alternatively, one may wish to use the lower confidence bound or lower quantile of a harm's probability to minimise unnecessary regulatory requirements, choosing to underestimate risk in the face of uncertainty.
There are countless other approaches.

\inlineheading{Takeaways}
In this section, we discuss a few of the ways that (i) defining probability and (ii) measuring probability create ambiguity. 
Clarifying the meaning of probability is important relative to the balancing framework in Section \ref{sec:dimensions}, as ambiguity may contribute to `risk hacking,' where AI providers and deployers opt for the interpretation that suits them. 
For instance, an AI provider or deployer may adopt a frequentist view, which incentivises them to intentionally limit the amount of data they collect since no observations of an event result in an estimated probability of 0 for all harms. If harms occur with probability 0, then the risk they pose would also be 0. 
This is particularly concerning given that the AI Act leaves it to providers and deployers to conduct risk assessments of their systems for classification (Art. 5 and Art. 6(3) AIA), and risk management (Art. 9 AIA). 

Yet we advise caution --- there is no one-size-fits-all interpretation of probability, and in the absence of a canonical interpretation,
\emph{requiring estimates under multiple interpretations provides a more holistic picture}.

\subsection{Component 2: Interpreting `Severity'} \label{sec:severity}
Based on the definition of risk and its link to fundamental rights, we can deduce the following: severity of harm is intended to capture the extent to which a system interferes with an individual's fundamental rights.
However, quantifying severity is difficult for two reasons: measurability and lack of clarity on the correct `unit' of severity.

\inlineheading{Measurability} The severity of harm is difficult to measure, especially with respect to fundamental rights. 
We highlight three reasons: 
intangibility, perception, and traceability. 

Many rights are intangible, such as the right to privacy (Art. 7 CFR), or the right to the protection of personal data (Art. 8 CFR), which makes identifying (let alone, measuring) interference with these rights challenging. 
Unlike physical injuries or financial losses, privacy harms do not always leave visible evidence. 
For example, when someone’s personal data is exposed or monitored without consent, the harm may lie in the loss of control over one’s own life, the fear of future misuse, or the chilling effect on personal choices. 

Even when rights are tangible, such as the right to property (Art.~17 CFR), quantifying the severity of harm is difficult because it depends on the perspective from which one measures severity. 
For example, a piece of property has different value to different parties; one could assess its economic value, but this may not capture its inherent value to its owner.
Courts may take \emph{perceived} harm into account, considering how it arises at the level of the affected individual, a particular group, or society as a whole \citep{ECtHR2023practical}.
Accordingly, \citet{malgieri2025assessing} acknowledge that interferences should be assessed using subjective perceptions, in addition to objective parameters and real-life adverse effects. 

A third challenge is the traceability of harms. 
That is, while one may be able to assign a severity of harm value to downstream outcomes, one must still determine how many times each outcome occurs. 
If an AI system impacts a significant number of downstream outcomes, then quantifying severity may involve tracing the impact of AI on millions, billions, or even trillions of individuals and outcomes. 

\inlineheading{Unit of Severity}
A second and more fundamental challenge is the unit by which severity is quantified and compared. 
Often, all factors contributing to severity are cast in monetary units, e.g., euros, as money is a shared construct that unifies analyses.

For certain fundamental rights, like the right to property (Art. 17 CFR) or the right to conduct business (Art. 16 CFR), using monetary units to quantify severity is feasible. 
However, this monetary perspective is alien to other fundamental rights, such as the freedom of expression (Art. 11 CFR) or broad legal principles like democracy or the rule of law.
Against this backdrop, the Commission has---in the context of the AI Act's Impact Assessment---stated that the impact on fundamental rights could not be quantified, at least not in monetary value \citep{commission2021impact}.

That monetary value is alien to some fundamental rights does not mean that quantification or formalisation is impossible.
Courts have proposed assigning monetary value in terms of damages to the rights to life (Art.~2~CFR), to the integrity of the person (Art. 3 CFR), and to data protection (Art. 8 CFR).\footnote{For instance, in \textit{Bindl v EU Commission}, Case T-354/22, the CJEU ordered the EU Commission to pay €400 in damages to an individual whose personal data had been unlawfully transferred to a third country.} 
Moreover, the CJEU case-law confirms that the measurement of interferences, i.e. its seriousness, is necessary to assess proportionality of limitations on rights \citep{malgieri2025assessing}.\footnote{Cf. CJEU, \textit{La Quadrature du Net}, Case C-511/18, C-512/18, and C-520/18, par. 131; and CJEU, Slagelse Almennyttige Boligselskab Afdeling Schackenborgvænge, Case C-417/23, par.~168.}

One way to get around assigning units is by turning to an \emph{ordinal} analysis. 
For instance, several works suggest formalising the balancing decision between competing fundamental rights by developing an ordinal scale for severity rather than assigning monetary units \citep{alexy_balancing_2003,zufall2023towards,araszkiewicz_tipping_2026}.
Similar approaches have been proposed in the context of fundamental rights impact assessment \citep{leslie2022human, unesco2023ethical, mantelero2024fundamental}. 
After all, much of economics is built on the concept that ordinal preferences (i.e., preferring outcome A to B) are underpinned by cardinal preferences (i.e., the actual values of outcomes A and B).

In the context of the AI Act, each time a court balances one fundamental right with another (for instance, deciding that the protection of personal data outweighs the right to conduct business in a specific case), 
the court is expressing an `ordering' of factors. 
By observing decisions across time (e.g., A is preferred to B, B is preferred to C, etc.),
one can tease apart how the court weighs the relative impact of different factors.
Furthermore, an example of a preference ordering in the AI Act is the list in Annex III of high-risk use cases and exceptions made for `low-impact conditions' and limited harm to health, safety, and fundamental rights (Art. 6(3) AIA). It reflects a legislative decision on which use cases are considered, for instance, high-risk. 
A number of studies have proposed different scale ratings (e.g., low, medium, high, very high distributed in a geometrical scale) and parameters or dimensions (e.g., gravity, temporal extent, culpability, magnitude) to properly frame the severity of impacts \citep{alexy_balancing_2003, mantelero2024fundamental, malgieri2025assessing, sass2026quantifying, novelli2024ai}.
This suggests that using a coarse scale in addition to ordinal preferences can help to quantify severity.

\inlineheading{Takeaways}
We discuss interpretations of severity in light of the balancing framework in Section \ref{sec:dimensions}, demonstrating that it can be challenging to quantify due to measurability and questions of units. In particular, many fundamental rights and protected values (such as democracy and the rule of law) cannot be assigned monetary value, creating a `units' issue.
However, we argue that one \emph{can still assess severity `ordinally': the comparative degree to which they interfere with the fundamental rights at stake} \citep{commission2025guideline, malgieri2025assessing}.

\subsection{Component 3: Interpreting `Combination'}
\label{sec:quant_combination}
Below, we consider three interpretations of the `combination' of probability and severity. Importantly, these interpretations reflect different priorities and constraints, and regulators may wish to combine them, when possible. 

\begin{figure}[t]
    \centering
    \begin{subfigure}{0.24\textwidth}
        \centering
        \includegraphics[width=\textwidth]{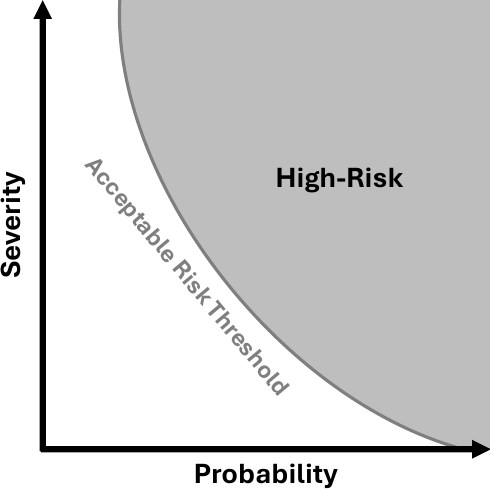}
        \caption{}
        \label{fig:exp_risk}
    \end{subfigure}
    \hfill
    \begin{subfigure}{0.24\textwidth}
        \centering
        \includegraphics[width=\textwidth]{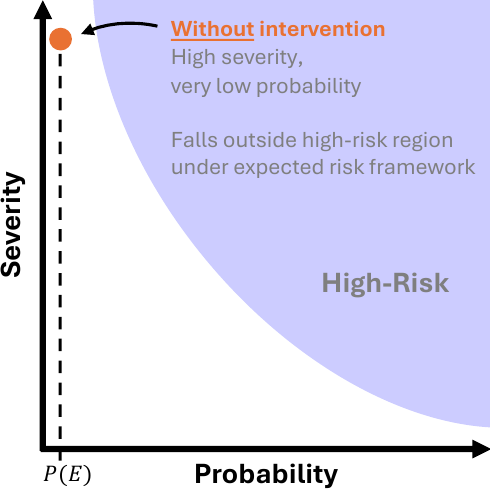}
        \caption{}
        \label{fig:exp_risk_gap}
    \end{subfigure}
    \hfill
    \begin{subfigure}{0.24\textwidth}
        \centering
        \includegraphics[width=\textwidth]{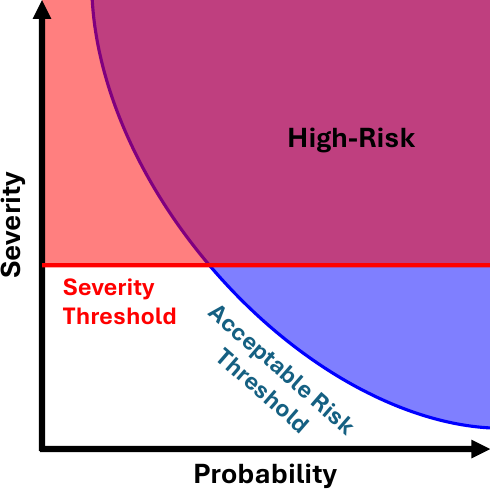}
        \caption{}
        \label{fig:sev_first}
    \end{subfigure}
    \hfill
    \begin{subfigure}{0.24\textwidth}
        \centering
        \includegraphics[width=\textwidth]{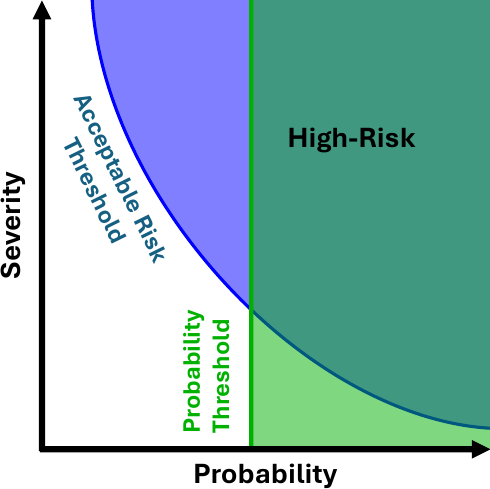}
        \caption{}
        \label{fig:prob_first}
    \end{subfigure}
    \caption{Illustrating different interpretations of `combinations' in the definition of risk. (a) Expected risk approach, (b) the `blind spot' of expected risk, (c) a severity-first approach, and (d) the analogous probability-first approach. }
    \label{fig:exp_risk_and_cost_benefit}
\end{figure}

\paragraph{Expected risk.}\label{sec:exp_risk}
The first and most straightforward interpretation of combination is expected risk. A well known concept in statistics and economics, expected risk is the product of probability and severity, summed over `events' (i.e., outcomes or harms).\footnote{For example, suppose there are only two events. If they occur with 30 percent and 70 percent probability, respectively, and the first event's severity of harm is -1,000 and the second event's severity of harm is $0$, then the expected risk is $0.3 \times (-1000) + 0.7 \times 0$.} Formally, expected risk is defined as:
$\text{Expected Risk} = \sum_{E \in \mathcal{E}} P(E) \times s(E)$,
where $\mathcal{E}$ is the set of all possible events (i.e., outcomes or harms), $E$ is a given event, $P(E)$ is the probability of event $E$, and $s(E)$ is the severity of the harm posed by event $E$. 

One can interpret expected risk as an \emph{average}, as it weights the severity of each outcome by the probability that the outcome occurs. 
We visualise a highly simplified version of the expected risk approach in Figure \ref{fig:exp_risk}, in which we oversimplify expected risk for visualisation purposes by assuming there is only \emph{one} event $E$ of interest and thus expected risk is given by $P(E) \times s(E)$.
In such cases, an AI system is high-risk if  $P(E) \times s(E) > \text{Threshold}$.

In many cases, expected risk does not match how decision makers assess and act on risk, suggesting that expected risk may not be the most suitable interpretation. 
Specifically, an approach based solely on expected risk exhibits a gap that some may object to: \emph{potentially extreme-severity harms that have \emph{low} probability may be classified as `minimal risk' when using an expected-risk framework because expected risk is the \emph{product} of probability and severity} (Figure \ref{fig:exp_risk_gap}). This might be problematic with respect to the list of prohibited AI practices in Art. 5 AIA (see Section \ref{sec:discussion}).

\inlineheading{Balancing Analysis}\label{sec:cost_benefit}
This discrepancy between a decision made based solely on expected risk and the `common sense' decision reflects the \emph{presence of other factors}.

For example, Section \ref{sec:balancing} considers several factors, including the risk of the AI system without regulatory intervention;
the relative benefits 
of the intended purpose or use case; and
the costs associated with the regulatory intervention on providers, deployers, and regulators. 
In some ways, the balancing analysis mirrors a cost-benefit type of analysis --- if the relative benefits (e.g., the risk reduction due to regulation) outweigh the costs (e.g., costs to AI providers and deployers as well as regulators), 
then the chosen regulatory intervention is suitable.
One can then assess this for all possible regulatory interventions, deciding which intervention achieves the desired benefit-to-cost.

This reflects a `common-sense' assessment: the regulatory intervention should be proportional to reduce risk below an acceptable threshold.  
This approach is consistent with the balancing calculation in the framework in Section \ref{sec:balancing}, and we further discuss it in Section \ref{sec:discussion}.

\inlineheading{Severity-First and Probability-First Risk}
The previous interpretations of `combination' treated probability and severity equally, but there are interpretations of `combination' that prioritise one over the other. 

For example, a regulator may take what we call a `severity-first' approach. 
Under this approach, the regulator first assesses the severity of each AI system, applies a threshold on severity, then allocates resources based on severity. 
For instance, since estimating probability is difficult, the regulator may wish to only invest resources into estimating probability for AI systems with sufficiently high severity, 
as visualised in Figure \ref{fig:sev_first}.
An analogous `probability-first' approach is shown in Figure \ref{fig:prob_first}.

An important implication of this approach is that \emph{order matters}: to see this, observe that the overlap between the red and blue region in Figure \ref{fig:sev_first} is different than the green and blue region in Figure \ref{fig:prob_first}. This means the AI systems classified as high-risk under a severity-first interpretation of `combination' would differ from those under a probability-first interpretation. 

\inlineheading{Takeaways}
We demonstrate that the \textit{`combination'} of probability and severity of risk constitutes a nuanced concept in itself, again relative tp the framework in Section \ref{sec:dimensions}. A straightforward interpretation of it is to compute the `expected risk'. However, we argue that a number of factors complicate this interpretation, including relative costs and benefits and the intended purpose. In addition, resource constraints may impose a severity- or probability-first approach to measuring risks.
\section{Synthesis} \label{sec:discussion}

Given the two-step framework presented in Section \ref{sec:balancing} and the quantitative interpretations discussed in Section \ref{sec:cost_benefit}, \textit{how can fundamental rights and risk quantification be compatible?}  

\inlineheading{Compatibility of Fundamental Rights Protection and Risk Quantification}
The AI Act's concept of risk oscillates between the definition of risk as a ‘combination of the probability of an occurrence of harm and the severity of that harm’ and a qualitative reference point as harms to ‘health, safety or fundamental rights’.
Legal literature has highlighted that the notion of risk, particularly with regard to harm to fundamental rights, encompasses a qualitative dimension, calling its potential quantification into question \citep{mahler2021between, yeung2022demystifying, malgieri2025assessing}. 

However, our analysis of the AI Act shows that fundamental rights and risk quantification are not only compatible, but also intertwined.
The goal of the AI Act is to ensure that an AI system poses a sufficiently low risk to fundamental rights (when balanced against its intended use or purpose). 
In some cases, ensuring that the risk is acceptable requires regulatory intervention.
Yet unless \emph{all} AI systems, their providers, and their deployers receive regulatory intervention, 
then there must be some classification to differentiate interventions, 
and the AI Act does so by appealing to risk quantification.

The AI Act approaches this by assigning a risk classification appropriate to the degree of intervention necessary to manage risks and thereby justifying the interference with the fundamental right(s) at stake \citep{degregorio2022european, fraser2024acceptable, gasiola2025rebuilding}. 
Assessing risk requires operationalising the Act's definition of risk as the `combination' of the `severity' and `probability' of harm to health, safety, and fundamental rights. 
We show in Section \ref{sec:quant_risk} that all three components of the definition of risk are overloaded concepts with multiple interpretations. 

\paragraph{Balancing Analysis}
Based on fundamental rights doctrine, the AI Act seems to interpret `combination' as a balancing analysis.

While the interpretation of `combination' as `expected risk' may appear to be a straightforward interpretation, it is insufficient to frame the AI Act's classification criteria.  
Assessing expected risk alone fails to acknowledge the relative benefits of AI systems associated with the intended purpose and the impacts of regulatory intervention on providers, deployers, and regulators. 
Notably, while the list of prohibited AI practices delineates a threshold for the unacceptable-risk level (Art. 5 AIA), exceptions to this list (as discussed in Section \ref{sec:balancing}) reveal that the purpose of those systems might justify a classification to a lower risk level \citep{gasiola2025rebuilding}, despite the expected risks. 

The `balancing analysis' approach integrates the calculation of expected risks with the relative benefits of the intended purpose or use.
Under this framework, one considers:
\begin{enumerate}[label=(\arabic*)]
    \item the expected risk of the AI system without regulatory intervention;
    \item the relative benefits of the intended purpose;
    \item the reduction in expected risk under regulatory intervention given regulatory intervention;
    \item the costs associated with the regulatory intervention on providers, deployers, and regulators. 
\end{enumerate}
Then, the idea behind the AI Act is to choose the lowest risk classification such that, altogether, the balanced risk is {less than or equal to an acceptable level}.

This methodology considers the costs of regulatory intervention on providers, deployers, and regulators to justify risk management strategies (avoidance, reduction or acceptance). In essence, this methodological approach mirrors the two-step framework, whereby these considerations can be structured into two pairs of balancing actions: potential harm to protected rights against intended purpose, and the resulting risk level (without regulation) against the impact of regulatory intervention to lower the risk level (compare Fig.\ref{fig:balancingrisks}).

\inlineheading{AI Act Implies a Severity-First Approach}
Finally, we argue that the AI Act implicitly utilises a severity-first mindset and suggest a combination of a balancing analysis and a severity-first approach to support the two-step framework. 

The connection to severity-first can be viewed as follows.
First, a purely expected-risk framework would contradict the classification criteria in the AI Act. If an impact on fundamental rights has very high severity (one could say `unacceptably high') but extremely low probability, then its expected risk will be mild or even low; yet various impacts are deemed `unacceptable' in the AI Act, regardless of their probabilistic weighting. For instance, Art. 5(1)(h) AIA prohibits, without prejudice to exceptions, the use of `real-time' remote biometric identification systems in publicly accessible spaces for law enforcement, which considers only severity of the impact on fundamental rights (e.g., privacy, data protection, or non-discrimination).
In instances where the AI Act calls for probabilistic weighting (e.g., Art. 5(1)(a) and (b) AIA), this can be interpreted as a secondary evaluation following the initial severity assessment.
Relatedly, the list in Annex III considers only use cases with high-severity impacts to classify in the high-risk level,  
suggesting the priority of severity considerations. In a subsequent step, if the AI system is then discovered to create interferences with low probability, then it may no longer be considered high-risk (Art. 6(3) AIA). Notably, the severity-first approach is also supported by the literature, which suggests that considerations of probability are generally considered to be of secondary importance when it comes to interferences \citep{malgieri2025assessing, council2024methodology}.

The severity-first approach is furthermore aligned with the understanding of fundamental rights as moral boundaries to the activity of public and private bodies \citep{yeung2022demystifying}. In particular, when the interference of an AI system affects the recognised core of a fundamental right, its `harm' always outweighs the system's purpose or use (step 1), and the resulting risk level outweighs the impact of regulatory intervention (step 2), e.g. justifying prohibition. Nevertheless, if the assessment yields a very low probability, considerations of probability may result in a lower risk level, although regulatory intervention may still be justified (high-risk/mandatory requirements, instead of unacceptable/prohibition).

\section{Conclusion}

Our contribution shows that the qualitative and quantitative dimensions of risk can be compatible, and that the methods and choices used to attain this alignment matter. 
If the methodology of risk quantification and its connection to fundamental rights is left to providers and deployers, there is potential for `risk hacking', which could lead to the underclassification of AI systems and regulatory shortcuts.
Furthermore, considering the dimension of fundamental rights standing behind the AI Act is not only legally relevant with respect to interferences and potential violations of those rights.
These considerations may also continue to guide the technical and implementation choices to be made for operationalising risk --- such as for the recent Digital Omnibus on AI\footnote{Regulation (EU) 2026/1744 of the European Parliament and of the Council of 8 July 2026 amending Regulations (EU) 2024/1689, (EU) 2018/1139 and (EU) 2023/1230 as regards the simplification of the implementation of harmonised rules on artificial intelligence.} to simplify the AI Act's application.
Within the proposed two-step framework, strengthened and informed by sound risk quantification methodology, the operationalised concept of risk can help to achieve a balance between the legitimate purpose of AI systems, their potential harms to fundamental rights and protected values, and the impact of regulatory intervention. 
In this manner, regulation may ensure innovation while preserving an acceptable level of risk to individuals and society.

\section*{Acknowledgments} \label{sec:acknowledgments}
The research was partially funded by the Topic Engineering Secure Systems of the Helmholtz Association (HGF). It was supported by KASTEL Security Research Labs.

\bibliographystyle{apalike}
\bibliography{aaai2026}

\end{document}